\documentstyle[12pt]{article}

\begin{document}

\newcommand{\beq}{\begin{equation}}
\newcommand{\eeq}{\end{equation}}
\def\pr{\partial}
\def\e{\epsilon}
\def\l{\lambda}
\def\D{\Delta}

\begin{titlepage}

\begin{center}

\hfill March 1996
\hfill hep-th/9603118

\vskip 1.8cm
{\bf String/$(D-5)$-brane Duality and $S$ Duality \\
 as Symmetries of Actions }

\vskip .8cm

Shijong Ryang

\vskip .8cm
{\em Department of Physics \\ Kyoto Prefectural University of Medicine \\
Taishogun, Kyoto 603, Japan}

\end{center}

\vskip 2.0cm

\begin{center} {\bf Abstract} \end{center}
We realize the string/$(D-5)$-brane duality on the action
level between the $T^{10-D}$-compactified heterotic string effective action
and the $(D-5)$-brane effective action in $D$ dimensions by managing a
Lagrange multiplier field. A dual dictionary is composed to be available
for the translation between the elementary or solitonic solutions of the
dual pair of actions. In the same way the $S$ duality is
 also reconstructed on the action level
as a double dualization for the $T^6$-compactified heterotic string
effective action.

\end{titlepage}

Recently, there has been much attention to the dynamics and symmetries of
string and $p$-brane in various dimensions \cite{1,2,3,4,5}.
The non-perturbative aspects of string and $p$-brane theories have been
understood by  various kinds of dualities. There is a web of
interconnections among many string and $p$-brane theories, which may
suggest the existence of a single unified theory. The string-string
duality in six dimensions implying that the heterotic string compactified
on $T^4$ is dual to the type IIA string compactified on $K3$ has
emerged as a symmetry of the equation of motion and not the action
\cite{6,7}. This duality relates the elementary singular solution of
the heterotic string theory to a solitonic non-singular solution of
the type IIA string theory and similarly the elementary type IIA
string is regarded as a solitonic non-singular solution of the heterotic
string theory. In seven dimensions the heterotic string compactified
on $T^3$ has been considered to be equivalent to the $D=11$
supergravity compactified on $K3$ \cite{4}. This string-membrane duality
is shown on the action level where the starting heterotic string
effective action is dualized into the effective $D=7$ action obtained
through $K3$ compactification of the $D=11$ supergravity, by
introducing a Lagrange multiplier $3$-form potential \cite{8}.
The dualities in five dimensions have been also studied
\cite{9,10}. The $D=11$ supergravity compactified on $K3\times T^2$ is
identified with the heterotic string compactified on $T^5$ \cite{10}.

On the other hand there are different kinds of symmetries for the string
theory such as the $S$ duality and the $T$ duality \cite{11}.
The $T$ duality is a manifest symmetry of the low-energy string
effective action, while the $S$ duality is a symmetry of the equation
of motion only. But by introducing suitable auxiliary fields the $S$ duality
has been manifestly realized on the action level in the same way as the $T$
duality, at a sacrifice of the manifest general coordinate invariance
\cite{12}. The general coordinate invariance, however, can
be recovered by eliminating the auxiliary fields through their equations
of motion.

In this letter we will present a general framework of the string/$(D-5)$-
brane duality on the action by introducing a Lagrange multiplier $(D-4)$-
form potential for the low-energy effective action describing the heterotic
string compactified on $T^{(10-D)}$. The structure of this duality will be
argued and compared with that of the string-string duality in ten dimensions
 between the heterotic string and the type I string \cite{13}.
Based on the same framework we will reformulate the $S$ duality in four
dimensions. We will see that our demonstration of the $S$ duality on the
action is different from the prescription of Schwartz and Sen \cite{12}
and preserves the general coordinate invariance manifestly.

We begin to consider the heterotic string theory compactified on
$T^{(10-D)}$, whose massless bosonic fields are the metric $g_{\mu\nu}$,
the antisymmetric tensor $B_{\mu\nu}$, the $U(1)$ gauge fields
$A_{\mu}^{a}, a= 1,\dots ,(36-2D)$, the scalar dilaton field $\Phi$ and
the $(36-2D)\times (36-2D)$ matrix-valued scalar moduli field $M$ obeying
$MLM^{T}=L, M^{T}=M$ with
\beq
L=\left(\begin{array}{cc} -I_{26-D} &  \\ & I_{10-D} \end{array}
\right),
\eeq
where $I_k$ denotes the $k\times k$ identity matrix. The low-energy effective
action for these fields is expressed as \cite{14}
\begin{eqnarray}
S_h &= &\int d^{D}x\sqrt{-g}e^{-\Phi}[ R+ \pr_{\mu}\Phi\pr^{\mu}\Phi +
\frac{1}{8}Tr(\pr_{\mu}ML\pr^{\mu}ML) \nonumber \\
  &  & - \frac{1}{2\cdot3!}H_{\mu\nu\rho}H^{\mu\nu\rho}
  - F^{a}_{\mu\nu}(LML)_{ab}F^{b\mu\nu}],
\end{eqnarray}
where $F^{a}_{\mu\nu}=\pr_{\mu}A^{a}_\nu - \pr_{\nu}A^{a}_{\mu}$ and the
$3$-form antisymmetric  field strength is given by $ H_{\mu\nu\rho}
= \pr_{\mu}B_{\nu\rho}- 2A^{a}_{\mu}L_{ab}F^{b}_{\nu\rho} + $ cyclic,
which includes the Chern-Simons term and then satisfies the modified
Bianchi identity
\beq
\pr_{[\mu}H_{\nu\rho\sigma]} = -3F^{a}_{[\mu\nu}L_{ab}F^{b}_{\rho\sigma]}.
\eeq
This action expressed in terms of the string sigma-model metric is changed
into one in terms of the Einstein canonical metric by the conformal
rescaling
\beq
g_{\mu\nu}=e^{2\Phi/(D-2)}g^{c}_{\mu\nu},
\eeq
so that we have
\begin{eqnarray}
S_{h}^{c}&=&\int d^{D}x\sqrt{-g^c}[ R^c - \frac{1}{D-2}(\pr\Phi)^2
+ \frac{1}{8}Tr(\pr ML\pr ML) \nonumber \\
 & & -\frac{1}{2\cdot 3!}e^{-4\Phi/(D-2)}H^2-e^{-2\Phi/(D-2)}F(LML)F] ,
\end{eqnarray}
where appropriate contractions are taken. In this stage we make a dual
transformation by imposing the modified Bianchi identity $(3)$ on $H$ with
a Lagrange multiplier $(D-4)$-form potential $B'$ and treating
$H_{\mu\nu\rho}$ as an independent field \cite{15}.
The Lagrange multiplier term added to the action $(5)$ is described by
\beq
\int d^Dx\alpha \e^{\mu_1\dots\mu_D}(\pr_{\mu_1}H_{\mu_2\mu_3\mu_4} +
3F^{a}_{\mu_1\mu_2}L_{ab}F^{b}_{\mu_3\mu_4}){B'}_{\mu_5\dots\mu_D},
\eeq
whose parameter $\alpha$ will be fixed. The equation of motion for
$H_{\mu\nu\rho}$ yields
\beq
H^{\mu_1\mu_2\mu_3}=-\frac{6\alpha}{D-3}(\sqrt{-g})^{-1}e^{4\Phi/(D-2)}
\e^{\mu_1\dots\mu_D}{H'}_{\mu_4\dots\mu_D},
\eeq
where $H'$ is the $(D-3)$-form field strength of $B'$ represented by
${H'}_{\mu_1\dots\mu_{D-3}}=(D-3)\pr_{[\mu_1}{B'}_{\mu_2\dots\mu_{D-3}]}$.
Substituting this solution back into the total action and choosing the
parameter  $\alpha = \pm 1/(3!(D-4)!)$ we obtain
\begin{eqnarray}
S_{d}^{c}&=&\int d^D x[ \sqrt{-g^c}(\;\; R^c- \frac{1}{D-2}(\pr\Phi)^2
+ \frac{1}{8}Tr(\pr ML\pr ML) \nonumber  \\
 & & - \frac{1}{2(D-3)!}e^{4\Phi/(D-2)}{H'}^2
 - e^{-2\Phi /(D-2)}F(LML)F\;\; ) + 3\alpha\e FLFB' ],
\end{eqnarray}
where in the last term the space-time indices are contracted by using the
completely antisymmetric $D$-dimensional $\e$ tensor. We have the form of
a dual action with a topological term, where the $(D-3)$-form
antisymmetric  field strength without the Chern-Simons term replaces
the $3$-form antisymmetric  field strength with the Chern-Simons
term. Here we make a minus choice  $\alpha=-1/(3!(D-4)!)$ and
change the sign of dilaton field as ${\Phi}'=-\Phi$. This sign flipped
action in terms of ${\Phi}'$ is denoted by ${S'}^c$.

From now on we restrict ourselves to $D>4$. Further the transformation by
the conformal rescaling
\beq
{g'}_{\mu\nu}=e^{4{\Phi}'/(D-2)(D-4)}g^{c}_{\mu\nu}
\eeq
provides an action in terms of the $(D-5)$-brane sigma-model metric
\begin{eqnarray}
S'&=&\int d^Dx[ \sqrt{-g'} \{ e^{-2{\Phi}'/(D-4)}( R' +
\frac{10-D}{(D-4)^2}(\pr{\Phi}')^2 +\frac{1}{8}Tr(\pr ML\pr ML)
 - \frac{H'^2}{2(D-3)!}  ) \nonumber \\
  &  &  -F(LML)F  \} - \frac{1}{2(D-4)!}\e FLFB' ] .
\end{eqnarray}
In this way the total dual transformation is accomplished
on the level of the action and not only the equation of motion.
The $(D-5)$-brane coupling constant given by $g'=e^{{\Phi}'/(D-4)}$
is related with the string coupling constant
 $g=e^{\Phi/2}$ as $g'=g^{-2/(D-4)}$.
We see that the coefficient of the $F(LML)F$ term has no dilaton dependence
in contrast with the ${H'}^2$ term. In $D=6$ dimensions the dual action $S'$
is just the low-energy effective action for the type IIA string
compactified on $K3$, whose coefficient of the dilaton kinetic term is
positive. This total dual transformation recovers the string-string
duality \cite{7}. The minus sign of the topological term
$-\e FLFB'/4$ in $(10)$ is linked with the plus sign of
$H = (\sqrt{-g})^{-1}e^{\Phi}\e H'$ . In $D$ dimensions the
total dual transformation is summarized by
\begin{eqnarray}
g_{\mu\nu}&=&e^{-2{\Phi}'/(D-4)}{g'}_{\mu\nu},\;\;\Phi =- {\Phi}', \nonumber
 \\  H^{\mu_1\mu_2\mu_3}&=&\frac{1}{(D-3)!}(\sqrt{-g})^{-1}e^{4\Phi/(D-2)}
\e^{\mu_1\dots\mu_D}{H'}_{\mu_4\dots\mu_D}
\end{eqnarray}
as a dual dictionary, whose first equation is provided
by combining $(4)$ and $(9)$.
Let us consider the case that we first carry out the dual transformation
on the heterotic string effective action $S_h$ to obtain an action,
which we represent by $S_{hd}$. Directly from $S_{hd}$ we cannot obtain the
desired dual action $S'$ with positive coefficient of the dilaton kinetic
term, no matter how we try to make both conformal
rescaling and dilaton sign flip. But this first dualized action $S_{hd}$
can be changed into the canonical one $S_{d}^{c}$ by the same conformal
rescaling as $(4)$, and so we can again arrive at $S'$ along the same
route. The dual action $S'$ in $D=7$ dimensions is almost the same one as
 obtained in Ref.$[8]$ except for the kinetic term of dilaton field, whose
coefficient is negative. This minus sign is due to the direct transformation
without going into the canonical actions as intermediates.

The dual transformation in the inverse direction can be also performed.
For instance in $D=6, 7$ dimensions we add to the respective actions
$S_{d}^{c}$ the Lagrange multiplier terms expressed by using  $2$-form
potentials $\hat{B}'$ as
\begin{eqnarray}
\int d^6x \frac{1}{2\cdot3!}\e^{\mu_1\dots\mu_6}\pr_{\mu_1}{H'}_
{\mu_2\mu_3 \mu_4}{\hat B'}_{\mu_5\mu_6}, \nonumber \\
\int d^7x \frac{1}{2\cdot4!}\e^{\mu_1\dots\mu_7}\pr_{\mu_1}{H'}_
{\mu_2\mu_3\mu_4\mu_5}{\hat B}'_{\mu_6\mu_7} ,
\end{eqnarray}
which impose the usual Bianchi identities on $H'$. Before proceeding
further the respective topological terms in $S_{d}^{c}$ are rewritten by
partial integrations as $\frac{1}{3!}\e ALFH', \frac{1}{4!}
\e ALFH'$.   Since the antisymmetric
tensor field $B'$ does not appear anywhere, we can promote $H'$ to the
status as an independent field. The variations with respect to $H'$ lead to
$H'= (\sqrt{-g^c})^{-1}e^{-\Phi}\e (ALF- \frac{1}{2}\pr{\hat B}'),
H'= (\sqrt{-g^c})^{-1}e^{-4\Phi/5}\e (ALF- \frac{1}{2}\pr{\hat B}')$.
The substitution of each expression back into $S_{d}^{c}$ reproduces
$S_{h}^{c}$.

Now we are concerned with the relationships between the elementary or
solitonic solutions in the above actions. We ristrict ourselves
to the neutral sector specified by setting the gauge fields to zero and begin
to write down an elementary solution with electric charge for the heterotic
string effective action $S_{h}^{c}$ in the canonical metric
\beq
ds_{h}^{c2} = \D^{-\frac{D-4}{D-2}}{\eta}_{\mu\nu}dx^{\mu}dx^{\nu} +
\D^{\frac{2}{D-2}}{\delta}_{mn}dy^m dy^n , \;\;\;
e^{-\Phi}= \D
\eeq
with $\D =1+ C_2/y^{D-4}$,
where $x^{\mu}(\mu =0, 1)$ are the coordinates of the string world surface
and $y^m$ are those of the $(D-2)$-dimensional transverse space and
$y= \sqrt{{\delta}_{mn} y^m y^n}$ \cite{16,17,18}.  Its electric charge
is proportional to the constant $C_2$.  Under the dual transformation this
electrically charged solution goes over to a magnetically charged one with
both metric and dilaton unchanged for the canonical interpolating
 action $S_{d}^{c}$.
Further by the dilaton sign flip and the succeeding rescaling of metric
$(9)$ we have a solitonic string solution for the $(D-5)$-brane effective
action $S'$
\beq
d{s'}^2  =  e^{4{\Phi}'/(D-2)(D-4)} ds_{h}^{c2}
         =  \D^{-\frac{D-6}{D-4}}\eta_{\mu\nu}dx^{\mu}dx^{\nu} +
 \D^{\frac{2}{D-4}}dy^m dy^m , \;\;\;
e^{{\Phi}'}= \D ,
\eeq
whose magnetic charge is specified by $C_2$. We see that in $D=6$ dimensions
the $x^{\mu}$ plane is flat. In the opposite direction the starting solution
$(13)$ is mapped to a solution for the heterotic string effective action
$S_h$ in the string sigma-model metric by the conformal rescaling $(4)$
\beq
ds_{h}^{2} = e^{2\Phi/(D-2)}ds_{h}^{c2}= \D^{-1}\eta_{\mu\nu}dx^{\mu}
dx^{\nu} + dy^m dy^m  , \;\;\; e^{-\Phi} = \D  .
\eeq
This electrically charged solution is translated into $(14)$ through the
dual dictionary $(11)$. Conversely as a starting solution we take an
elementary $(D-5)$-brane solution for the  $(D-5)$-brane effective
canonical action ${S'}^c$
\begin{eqnarray}
d{s'}_{c}^{2}&=& {\D'}^{-\frac{2}{D-2}}\eta_{\mu\nu}dx^{\mu}dx^{\nu} +
 {\D'}^{\frac{D-4}{D-2}}dy^m dy^m , \nonumber \\
 e^{-\Phi'}& =& \D'  , \;\;\; \D'= 1+ \frac{C_{D-4}}{y^2}
\end{eqnarray}
with electric charge proportional to $C_{D-4}$. The coordinates of the
$(D-5)$-brane world volume and those of the four-dimensional transverse
space are denoted by $x^{\mu}(\mu=0\dots(D-5)), y^m$ respectively.
We will trace back on the inverse route. Through the dilaton sign flip and
the previously discussed inverse dual transformation it becomes a solitonic
$(D-5)$-brane solution with magnetic charge for the string effective
canonical action $S_{h}^{c}$. Further the rescaling of metric yields a
solitonic solution for $S_h$
\beq
ds_{h}^{2}= e^{2\Phi/(D-2)}d{s'_c}^2= \eta_{\mu\nu}dx^{\mu}dx^{\nu}
 + \D'dy^mdy^m,\;\;\;  e^{\Phi} = \D'
\eeq
with magnetic charge proportional to $C_{D-4}$. From $(16)$ an elementary
$(D-5)$-brane solution with electric charge for $S'$ is provided through
 the conformal rescaling by
\beq
d{s'}^2= e^{4\Phi'/(D-2)(D-4)}d{s'}_{c}^{2} = \D'^{-\frac{2}{D-4}}
\eta_{\mu\nu}dx^{\mu}dx^{\nu} + \D'^{\frac{D-6}{D-4}}dy^mdy^m
\eeq
with $\Phi'$ left intact. In $D=6$ dimensions the solutions $(14), (18)$
turn out to be of the same forms as $(17), (15)$ respectively \cite{7}.
The pair of solutions $(17), (18)$ are also related with each other through
the dual dictionary $(11)$. Thus the dual dictionary should be applied to
the solutions of equations of motion but when used for the transformations of
the relevant actions it should be decomposed as previously argued.

Moreover we proceed to the singularity structure of these solutions.
In terms of the Schwarzschild-like coordinate $r$ defined by
$y^{D-4}= r^{D-4} -a^{D-4}$ with $C_2= a^{D-4}$ we rewrite $(15)$ and $(14)$
as
\begin{eqnarray}
ds_{h}^{2}&=& Adx^2 + A^{-\frac{2(D-5)}{D-4}}dr^2 + A^{\frac{2}{D-4}}r^2
d\Omega_{D-3}^{2} , \nonumber \\
d{s'}^2&=& A^{\frac{D-6}{D-4}}dx^2 + A^{-2}dr^2 + r^2d\Omega_{D-3}^{2} ,
\end{eqnarray}
where $A= 1- (a/r)^{D-4}$ and $d\Omega_{D-3}^{2}$ is the standard metric
on the unit $(D-3)$-sphere. In order to see the singularity structure
near $r=a$ we reparametrize $r$ as $(D-4)(r-a)/a=e^{(D-4)\rho/a}$ and
obtain in the asymptotic limit $r \rightarrow a, \rho
\rightarrow -\infty$
\begin{eqnarray}
ds_{h}^{2}&\sim& e^{(D-4)\rho/a}dx^2 + e^{2\rho/a}(d\rho^2 +
a^2d\Omega_{D-3}^{2}) , \nonumber \\
d{s'}^2 &\sim& e^{(D-6)\rho/a}dx^2 + d\rho^2 + a^2d\Omega_{D-3}^{2}
\end{eqnarray}
together with the linear dilatons $\Phi= -\Phi' \sim(D-4)\rho/a$,
which show that the weak heterotic string coupling limit is identified
with the strong $(D-5)$-brane coupling limit for the solitonic string
solution. The asymptotic metric for $ds_{h}^{2}$ is a warped product of
two-dimensional Minkowski space with the conformal factor shrinked for
$D>4$ and a cylindrical throat $R\times S^{D-3}$. This solution is
singular at the horizon $r=a$ and compared with the non-singular solution
$d{s'}^2$, which is a product of $S^{D-3}$ and three-dimensional anti-de
Sitter space with the conformal factor blown up for $D<6$ or shrinked for
$D>6$. In the critical $D=6$ dimensions it becomes a product of
three-dimensional Minkowski space and $S^3$ \cite{7}. On the other hand
by making the similar change of variable as $y^2=r^2-a^2$ with
$C_{D-4}=a^2$ for $(17)$ and $(18)$ we have
\begin{eqnarray}
ds_{h}^{2}&=& dx^2 + B^{-2}dr^2 +r^2d\Omega_{3}^{2} , \nonumber \\
d{s'}^2& =& B^{\frac{2}{D-4}}(dx^2 + B^{-2}dr^2 + r^2d\Omega_{3}^{2})
\end{eqnarray}
with $B=1- (a/r)^2$. In terms of $\rho$ defined by $e^{2\rho/a}=2(r-a)/a$
the asymptotic solutions are obtained by
\begin{eqnarray}
ds_{h}^{2} &\sim& dx^2 + d\rho^2 + a^2d\Omega_{3}^{2} , \nonumber \\
d{s'}^2 &\sim& e^{4\rho/(D-4)a}(dx^2 + d\rho^2 + a^2d\Omega_{3}^{2})
\end{eqnarray}
together with the linear dilatons $\Phi=-{\Phi}' \sim -2\rho/a$. Thus
the non-singular solitonic $(D-5)$-brane solution in the strong coupling
region for the heterotic string  theory is described by a weakly coupled
singular elementary $(D-5)$-brane solution for the $(D-5)$-brane theory.

In a similar viewpoint  we would like to comment on the
string-string duality in ten dimensions relating the heterotic string
theory to the type I string theory \cite{13}.  We start with the
ten-dimensional heterotic string effective action
\beq
S_h = \int d^{10}x\sqrt{-g}e^{-2\phi}[ R+ 4(\pr\phi)^2 -
\frac{1}{12}H^2 - F^2 ],
\eeq
which is similar to $(2)$ with $\Phi$ replaced by $2\phi$ and use the
conformal rescaling $g_{\mu\nu} =e^{\phi/2}g^{c}_{\mu\nu},(4)$ with
$D=10$ to obtain
\beq
S_{h}^{c}=\int d^{10}x\sqrt{-g^c}[R^c -\frac{1}{2}(\pr\phi)^2 -
\frac{1}{12}e^{-\phi}H^2 - e^{-\phi/2}F^2 ],
\eeq
which is read off from $(5)$. Skipping the dual transformation we change
the sign of dilaton field as $\phi'=-\phi$ to get ${S'}^c$. The conformal
rescaling ${g'}_{\mu\nu}=e^{\phi'/2}g^{c}_{\mu\nu}$ for ${S'}^c$ yields
\beq
S'=\int d^{10}x\sqrt{-g'}[e^{-2\phi'}( R+4(\pr\phi')^2) - \frac{1}{12}H^2
-e^{-\phi'}F^2 ],
\eeq
which is the low-energy effective action for the type I string.
The total transformation composed of these three steps is summarized by
$g_{\mu\nu}=e^{-\phi'}g'_{\mu\nu}
, \phi=- \phi'$  as a dual dictionary. This
string-string duality includes the strong-weak coupling exchange and
is compared with the previous string/$(D-5)$-brane duality which
contains it together with the electric-magnetic charge exchange.

Here we take up the $D=4$ case separately to consider the $S$
duality. By choosing plus sign as $\alpha =1/3!$ for $S_{d}^{c}, (8)$
and expressing the $0$-form potential $B'$ by $\psi$ we obtain
\begin{eqnarray}
S_{d}^{c} & = & \int d^4x\sqrt{-g^c} [ R^c - \frac{1}{2(\l_2)^2}
\pr \l \pr \bar{\l} + \frac{1}{8}Tr(\pr ML\pr ML)  \nonumber \\
  &  & - {\l}_2F(LML)F + {\l}_1FL \tilde{F} ],
\end{eqnarray}
where $\l=\l_1 + i\l_2 = \psi + ie^{-\Phi}$ and $\tilde{F}^{a\mu\nu} =
\frac{1}{2}(\sqrt{-g^c})^{-1}\e^{\mu\nu\rho\sigma}F^{a}_{\rho\sigma}$.
The plus sign of the last topological term is connected with the minus
sign of $H^{\mu\nu\rho}=-(\sqrt{-g^c})^{-1}e^{2\Phi}\e^{\mu\nu\rho\sigma}
\pr_{\sigma}\psi$. This sign choice is opposite to the previous one.
With respect to the $U(1)$ gauge fields we dualize this action by adding
$\int d^4x 2\beta\e ^{\mu\nu\rho\sigma}F^{a}_{\mu\nu}\pr_{\rho}
D_{\sigma}^{a}$ to ensure the usual Bianchi identities, where
$D_{\sigma}^{a}$ are Lagrange multiplier $1$-form potentials. The variations
over $F$ yield
\beq
C^{a}_{\mu\nu}=-\frac{1}{\beta}\left( \l_2(LML)_{ab}\tilde{F}^{b}_{\mu\nu}
+ \l_1L_{ab}F^{b}_{\mu\nu} \right),
\eeq
where $C^{a}_{\mu\nu}$ are the field strengths of $D_{\mu}^{a},
C^{a}_{\mu\nu}= \pr_{\mu}D_{\nu}^{a} - \pr_{\nu}D_{\mu}^{a}$. Before
proceeding by using the self-dual and anti-self-dual gauge field strengths
$F_{\pm}=-MLF \pm i\tilde{F}$ satisfying $\tilde{F}_{\pm} = \pm iMLF_{\pm}$
we rewrite the sum of the kinetic  term and the topological term in
$(26)$ as
\beq
\frac{i}{4} (\l F_{+}M^{-1}F_{+} - \bar{\l}F_{-}M^{-1}F_{-} ),
\eeq
where we have taken account of $F_{+}M^{-1}F_{-} =0$, that holds
 due to $M^{T}=M$. Combining $(27)$ with its Hodge dual to give
\begin{eqnarray}
F=\frac{\beta ML}{2}\left( \frac{F_{+}^{c}}{\l} + \frac{F_{-}^{c}}
{\bar{\l}} \right) &,& \tilde{F} = \frac{i\beta}{2} \left( \frac
{F_{+}^{c}}{\l} -\frac{F_{-}^{c}}{\bar{\l}} \right),
\end{eqnarray}
where $F^{c}_{\pm} = -MC \pm iL\tilde{C}$ satisfying ${\tilde{F}}_{\pm}^{c}
=\pm iMLF_{\pm}^{c}$ and substituting them back into
the total action we obtain for the associated three terms
\beq
-\frac{i\beta^2}{4} \left( \frac{1}{\l}F_{+}^{c}M^{-1}F_{+}^{c} -
\frac{1}{\bar{\l}}F_{-}^{c}M^{-1}F_{-}^{c} \right) .
\eeq
The transformed self-dual and anti-self-dual gauge
field strengths are alternatively expressed as
$F^{c}_{\pm}=-\frac{1}{\beta}(\l_1 \pm i\l_2)F_{\pm}$. We find that if
we choose $\beta =1$ and set $\l' =- 1/\l$ the obtained compact expression
$(30)$ is of the same form as the initial sum $(28)$.
 Since the kinetic term of the  complex field
$\l$ is also rewritten by $-\pr \l'\bar {\pr} \l'/2({\l'}_2)^2$, the
transformed action becomes the initial one, but with
\begin{eqnarray}
{\l}'=-\frac{1}{\l},& {F'}_{+}=-{\l}F_{+} ,& {F'}_{-}=-\bar {\l}F_{-},
\nonumber \\ {M}'=M ,& {g'}^{c}_{\mu\nu}=g^{c}_{\mu\nu} . &
\end{eqnarray}
The action remains of the same form under the other transformation
$\l \rightarrow \l+c$ with a real constant $c$, that is combined with the
above dual transformation to generate the $S$ duality group $SL(2,R)$.

In conclusion we have elucidated the structure of the general
string/$(D-5)$-brane duality on the action level between the heterotic
string effective action in terms of the string sigma-model metric and
the $(D-5)$-brane effective action in terms of the
$(D-5)$-brane sigma-model metric in $D$ dimensions.
By taking advantage of  Lagrange multiplier fields we have presented
the dual as well as inverse dual transformations and seen that
the Chern-Simons term in the $3$-form field strength for the heterotic
string theory plays  an important role to give the topological term
in the dualized $(D-5)$-brane effective action. We have observed that
the string/$(D-5)$-brane duality is constructed from the four steps.
In order to produce the full dual transformation on the action level
by arranging the four steps it is essential that the relevant actions
in terms of the canonical metric interpolate in the route of
 transformations. We would like to emphasize that the dual dictionary
 itself should be used for the translation between the elementary or
solitonic solutions of equations of motion for the dual pair of actions.
The ten-dimensional string-string duality associated with the heterotic
and type I strings can be interpreted to be included within the larger
duality, the string/$(D-5)$-brane duality. The former duality consists of
the strong-weak coupling exchange and two decomposed conformal rescalings,
while the latter wide class of duality consists of them as well as the
electric-magnetic charge exchange. The $S$ duality in $D=4$ dimensions
has been realized on the action level through the double dual transformation
with respect to the gauge fields and the antisymmetric tensor field.
Both $S$ duality and string/$(D-5)$-brane duality can be understood
on an equal footing.
We hope that our study will give  a step toward classifying various
kinds of dualities.

\end{document}